\documentclass[9pt,twocolumn,twoside]{pnas-new}

\templatetype{pnasresearcharticle} 

\title{\huge Quasi-Diabatic Scheme for Non-adiabatic On-the-fly Simulations}

\author[a,b]{Wanghuai Zhou}
\author[b]{Arkajit Mandal} 
\author[b,1]{Pengfei Huo}

\affil[a]{School of Science, and Advanced Functional Material and Photoelectric Technology Research Institution,
Hubei University of Automotive Technology, Shiyan, Hubei, 442002, People's Republic of China,}
\affil[b]{Department of Chemistry, University of Rochester, 120 Trustee Road, Rochester, New York 14627, United States}

\leadauthor{Zhou} 

\significancestatement{This paper provides the first {\it ab-initio} on-the-fly example of using the Quasi-Diabatic (QD) scheme for non-adiabatic simulations with diabatic dynamics approaches. The QD scheme provides a seamless interface between {\it diabatic} quantum dynamics approaches and {\it adiabatic} electronic structure calculations. It completely avoids additional theoretical efforts to reformulate the equation of motion from diabatic to adiabatic representation, or construct global diabatic surfaces. This scheme enables many recently developed diabatic quantum dynamics approaches for {\it ab-inito} on-the-fly simulations, providing the non-adiabatic community a wide variety of approaches (such as the real-time path integral method and symmetric quasi-classical approach) beyond the well-explored methods (like trajectory surface-hopping or {\it ab-initio} multiple-spawning). The QD scheme also enables using realistic test cases (like ethylene photodynamics) that go beyond simple model systems to assess the accuracy and limitation of recently developed quantum dynamics approaches.}

\authorcontributions{W.Z, A. M. and P.H. designed research; W.Z.and A.M. performed research; W.Z., A.M. and P.H. analyzed data; and W.Z., A.M. and P.H. wrote the paper.}
\authordeclaration{The authors declare no conflict of interest.}
\correspondingauthor{\textsuperscript{1}To whom correspondence should be addressed. E-mail: pengfei.huo@rochester.edu}


\begin{abstract}
We use the quasi-diabatic (QD) propagation scheme to perform on-the-fly non-adiabatic simulations of the photodynamics of ethylene. The QD scheme enables a seamless interface between accurate diabatic-based quantum dynamics approaches and adiabatic electronic structure calculations, explicitly avoiding any efforts to construct global diabatic states or reformulate the diabatic dynamics approach to the adiabatic representation. Using partial linearized path-integral approach and symmetrical quasi-classical approach as the diabatic dynamics methods, the QD propagation scheme enables direct non-adiabatic simulation with the CASSCF on-the-fly electronic structure calculations. The population dynamics obtained from both approaches are in a close agreement with the quantum wavepacket based method and outperform the widely used trajectory surface hopping approach. Further analysis on the ethylene photo-deactivation pathways demonstrates the correct predictions of competing processes of non-radiative relaxation mechanism through various conical intersections. This work provides the foundation of using accurate diabatic dynamics approaches and on-the-fly adiabatic electronic structure information to perform ab-initio non-adiabatic simulation.\end{abstract}

\dates{This manuscript was compiled on \today}

\begin{document}

\maketitle
\thispagestyle{firststyle}
\ifthenelse{\boolean{shortarticle}}{\ifthenelse{\boolean{singlecolumn}}{\abscontentformatted}{\abscontent}}{}


\dropcap{N}onadiabatic Molecular Dynamics (NAMD) simulation plays an indispensable role in investigating photochemical and photophysical processes of molecular systems~\cite{TullyNAMD, SubotnikDIA, Levine2007arpcl,Nelson2014,Huo2016ARPC,Barbatti2018ChemRev,Curchod2018chemrev}. The essentially task of NAMD~\cite{TullyNAMD} is to solve the coupled electronic-nuclear dynamics governed by the total Hamiltonian of the molecular system, $\hat {H} = \hat {T} + \hat{V}(r,R)$, where $r$ and $R$ represent the electronic and nuclear degrees of freedom (DOF), respectively, $\hat {T}=-\frac{\hbar^2}{2M}\nabla^2_{R}$ is the nuclear kinetic operator, and $\hat{V}({r}, {R})$ is the electronic potential that describes the kinetic energy of electrons and electron-electron as well as electronic-nuclear interactions. Rather than directly solving the time-dependent Schr$\ddot{\mathrm o}$dinger equation (TDSE) governed by $\hat{H}$, NAMD simulation is usually accomplished~\cite{TullyNAMD,SubotnikDIA} by performing on-the-fly electronic structure calculations that provide the energy and gradients, and the quantum dynamics simulations that propagate the motion of the nuclear DOF (described by trajectories or nuclear wavefunctions). In particular, the electronic structure calculations solve the following eigenequation 
\begin{equation}
\hat {V} (r,R)|\Phi_{\alpha}(R)\rangle = E_{\alpha}(R)|\Phi_{\alpha}(R)\rangle,
\end{equation}
provides the {\it adiabatic} state $|\Phi_{\alpha}(R)\rangle$ and energy $E_{\alpha}(R)$. 

Because of the readily available electronic structure information in the {\it adiabatic} representation, quantum dynamics approaches formulated in this representation have been extensively used to perform on-the-fly NAMD simulations, including the popular fewest-switches surface hopping (FSSH)~\cite{Tully,Rossky-Webster,space1991nonadiabatic,tully94jcp,subotnik2016arpc,Nelson2014,Jain2016JCTC,PrezdoSH,BarbattiSH}, {\it ab-initio} multiple spawning (AIMS)~\cite{Levine2007arpcl,Ben-Nun2000aims,Curchod2018chemrev}, and several recently developed Gaussian wavepacket approaches~\cite{mcEhrenfest,DmitryJCP2014,FernandezPCCP2015,Izmaylov2018JCP,Meek2016JCP}, coupled-trajectory approaches~\cite{Gross17,Curchod18,Rubio18,Martens}, and the {\it ab-initio} multi-configuration time-dependent Hartree (MCTDH)~\cite{Habershon2018,Habershon2018JCTC}. Among them, FSSH is one of the most popular approaches in NAMD, which uses mixed quantum-classical (MQC) treatment of the electronic and nuclear DOFs that provides efficient non-adiabatic simulation. As a MQC method, however, FSSH treats quantum and classical DOF on different footings~\cite{TullyNAMD}, generating artificial electronic coherence~\cite{Tully,subotnik2016arpc} that give rise to incorrect chemical kinetics~\cite{subotnik2016arpc} or the breakdown of the detailed balance (time-reversibility)~\cite{Schmidt:2008}. Recently developed non-adiabatic quantum dynamics approach~\cite{Kelly:2012,HuoPLDM, Hsieh:2013,Walters2016JCP,Ananth:2013,Richardson:2013,Chowdhury:2017,SCFaradayMiller,WHMiller2015jcp,Geva2019,Richardson2019} have shown a great promise to address the deficiency and limitations of MQC approximation. However, these approaches are usually developed in the {\it diabatic} representation and incompatible with the available adiabatic electronic structure calculations. Reformulating them back to the adiabatic representation requires additional and sometimes, non-trivial theoretical efforts. 

To address this discrepancy, we have developed the Quasi-Diabatic (QD) propagation scheme~\cite{Mandal:2018,Sandoval2018} which provides a seamless interface between accurate {\it diabatic} quantum dynamics approaches and routinely available {\it adiabatic} electronic structure information for on-the-fly simulations. The {\it key conceptual breakthrough} behind the QD scheme is by recognizing that, in order to propagate quantum dynamics with diabatic dynamics approaches, one only needs {\it locally} well-defined diabatic states, as oppose to construct {\it global} dibabatic states from the diabatization procedures~\cite{Baer:1976,Mead:1982,Cave:1997, Subotnik:2008, SubotnikDIA, DiabaticARPC,Yarkony:2016}. These local diabatic states can simply be adiabatic states with a reference geometry, which are commonly referred as the crude adiabatic states. Consider a short-time propagation of the nuclear DOFs during $t\in[t_0, t_1]$, where the nuclear positions evolve from ${R}(t_0)$ to ${R}(t_1)$, and the corresponding adiabatic states are $\{|\Phi_{\alpha}({R}(t_0))\rangle\}$ and $\{|\Phi_{\lambda}(R(t_1))\rangle\}$. The QD scheme uses the nuclear geometry at time $t_0$ as a reference geometry, ${R_{0}}\equiv {R}(t_0)$, and use the adiabatic basis $\{|\Phi_{\alpha}({R}(t_0))\rangle\}$ as the {\it quasi-diabatic} basis during this short-time propagation, such that
\begin{equation}\label{eqn:qdidea}
|\Phi_{\alpha}({R_{0}})\rangle\equiv|\Phi_{\alpha}({R}(t_0))\rangle,~~\mathrm{for}~t\in[t_0,t_1].
\end{equation}
With the above QD basis, derivative couplings vanish during each propagation segment, and at the same time, $\hat{V}({R})$ has off-diagonal elements in contrast to the pure diagonal matrix under the adiabatic representation. With this local diabatic basis, all of the necessary diabatic quantities can be evaluated and used to propagate quantum dynamics during $t\in[t_0,t_1]$. During the next short-time propagation segment $t\in[t_1,t_2]$, the new QD basis $|\Phi_{\lambda}({R'_{0}})\rangle\equiv|\Phi_{\lambda}({R}(t_1))\rangle$ will be used to propagate the quantum dynamics, and any state-dependent quantities will be transformed from the $|\Phi_{\alpha}({R_{0}})\rangle$ basis to the $|\Phi_{\lambda}'({R'_{0}})\rangle$ basis. With the nuclear geometry closely following the reference geometry at every single propagation step, the QD basis forms a convenient and compact basis in each short-time propagation segment. To summarize, the QD scheme~\cite{Mandal:2018,Sandoval2018,Mandal2018jcp,Mandal2019jpca} uses the adiabatic states associated with a reference geometry as the quasi-diabatic states during a short-time quantum propagation, and dynamically update the definition of the QD states along the time-dependent nuclear trajectory. It allows a seamless interface between diabatic dynamics approaches with adiabatic electronic structure calculations. It also enables using realistic {\it ab-initio} test cases to assess the accuracy and limitation of recently developed quantum dynamics approaches~\cite{Mandal2019jpca}.

In this paper, we provide the first {\it ab-initio} on-the-fly example of using the QD scheme~\cite{Mandal:2018} for non-adiabatic simulations with {\it diabatic} quantum dynamics approach and the adiabatic electronic structure calculations. In particular, we use two recently developed diabatic dynamics approaches, partial linearized density matrix (PLDM) path-integral method~\cite{HuoPLDM,Huo2016ARPC} and symmetric quasi-classical (SQC) window approach~\cite{MillerJCP13} to directly perform on-the-fly NAMD simulations of the well-studied ethylene photodynamics. On-the-fly electronic structures calculations are performed at the level of complete active space self-consistent field (CASSCF) approach. The results obtained from QD-PLDM and QD-SQC are in a close agreement with {\it ab-initio} multiple spawning (AIMS). Thus, this paper provides the first on-the-fly example of the QD propagation scheme, as well as completes the establishment of it in the field of {\it ab-initio} non-adiabatic dynamics as a powerful tool to enable accurate diabatic quantum dynamics approaches for on-the-fly simulations. 
\begin{figure}
\centering
\includegraphics[width=0.8\linewidth]{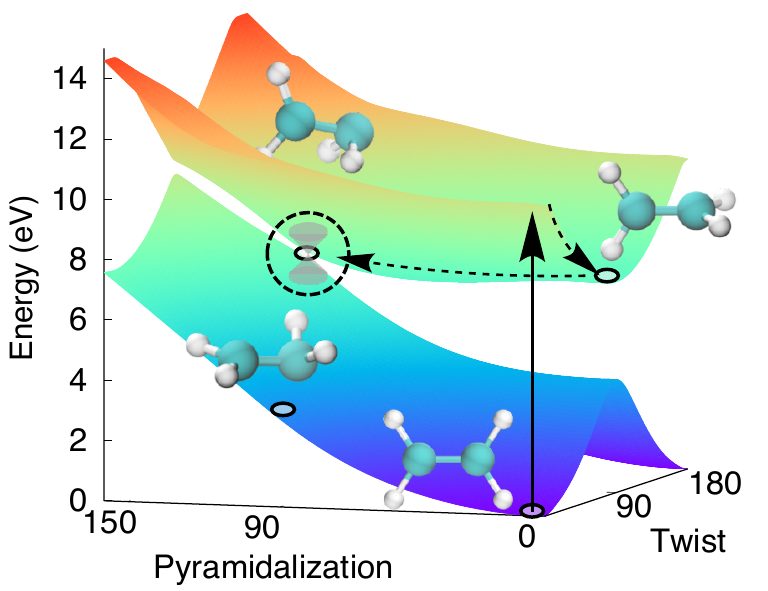}
\caption{CASSCF potential energy surface of $S_1$ (upper surface) and $S_0$ (lower surface) along two main reaction coordinates of photo-dissociation pathways in ethylene. Upon photo-excitation, the system quickly relax to the minimum along the twist angle on the $S_1$ surface, then pyramidalized and relax back to the $S_0$ through the twisted-pyramidalized conical intersection.}
\label{fig-pes}
\end{figure}

\section*{Results and Discussions}
Despite being one of the simplest conjugated molecules, ethylene exhibits a complex photo-dissociation dynamics by visiting several conical intersections and undergoing various reaction pathways during the non-radiative decay processes. It is thus considered as a prototype for investigating photo-isomerization reactions through conical intersections~\cite{Levine2007arpcl}, and has been extensively studied through theoretical~\cite{Ben-Nun1998cpl,Barbatti2004jcp,Barbatti2005cpl,Levine2008,Tao2009jpca,Mori2011} and experimental~\cite{Kosma2008,Kobayashi2015,Tao2011jcp,Allison2012jcp} investigations. To provide an accurate description of the electronic structure of ethylene, we follow the previous theoretical studies~\cite{Levine2008,Hollas2018} and use CASSCF approach that has shown to provide accurate potential around conical intersections. To avoid the root-flipping problem~\cite{Levine2008,Tao2009jpca,Hollas2018}, here, the CASSCF calculations are performed using state-averaging over three states, at the level of SA-3-CASSCF(2,2) with 6-31G* basis set, as implemented in MOLPRO~\cite{MOLPRO-WIREs}. The non-adiabatic dynamics simulation, is propagated in the $\{|\mathrm{S}_{0} (R)\rangle ,|\mathrm{S}_{1}(R)\rangle\}$ electronic states subspace, {\it i.e.}, the ground and the first excited states, by using the information from the on-the-fly CASSCF calculations. All of the QD-PLDM and QD-SQC approaches are implemented in a modified version of NAMD interface code SHARC~\cite{SHARC, SHARC2}, which are used to perform all of the simulations in this paper.

Fig.~\ref{fig-pes} presents the adiabatic potential energy surface (PES) of ethylene, with both $\mathrm{S}_1$ state (upper surface) and $\mathrm{S}_0$ state (lower surface) along the pyramidalization and the twist reaction coordinates, obtained from PES scans. The conical intersection among these two surfaces are also indicated with a dotted circle, located at a twist angle of $90^{\circ}$ and the pyramidalization angle around $108^{\circ}$. Upon the photoexcitation (indicated by the solid arrow), ethylene first relaxes on the $\mathrm{S}_1$ surface along the twist angle (indicated by the dash arrow), then pyramidalize on the $\mathrm{S}_1$ surface and reach to the region of the conical intersection (which is commonly referred as the twisted-pyramidalized conical intersection), and quickly relaxes back to the $S_0$ surface. This of course, is only a very simplified picture. The actual non-adiabatic dynamics is much more complex and a direct on-the-fly NAMD simulation is often necessary to reveal the fundamental mechanistic insights into these complex reaction channels~\cite{Levine2007arpcl,Barbatti2005cpl,Barbatti2005jcp,Tao2009jpca,Tao2011jcp}.

\begin{figure}
\centering
\includegraphics[width=1.0\linewidth]{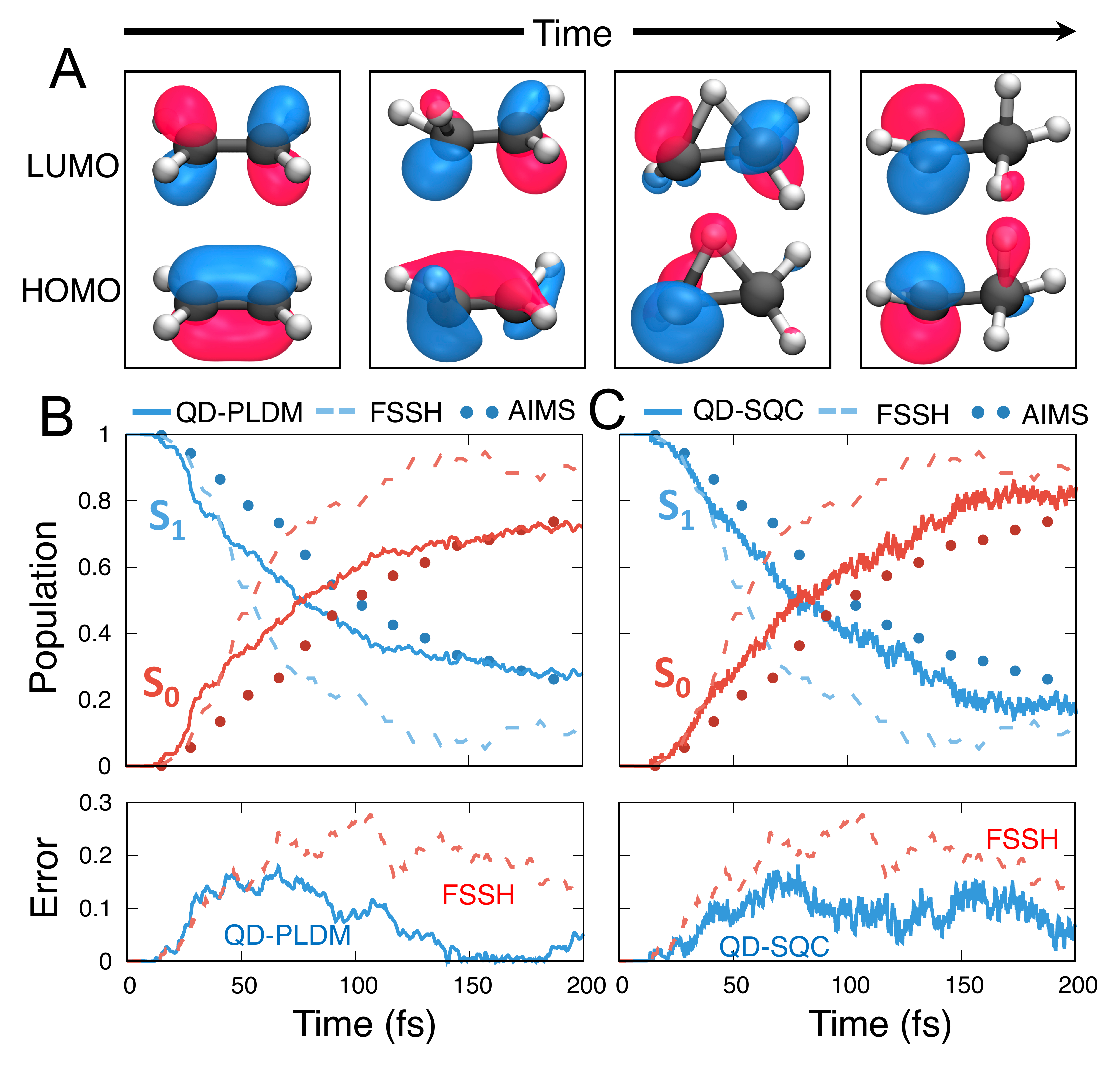}
\caption{Population dynamics obtained from the QD propagation scheme. A. The frontier molecular orbitals (HOMO and LUMO) along a given nuclear trajectory. B. The adiabatic electronic populations of S$_1$ (blue) and S$_0$ (red) obtained from QD-PLDM (solid lines), FSSH (dashed lines) and AIMS (filled circles), respectively. AIMS, which is an approximate Gaussian wavepacket-based non-adiabatic method, should be viewed as the benchmark result that is close to the numerically exact solution of the quantum dynamics in this CAS model. The bottom panel presents the time-dependent error between the trajectory-based approach and AIMS. C. The adiabatic electronic populations obtained from QD-SQC (solid lines), with the rest same as presented in B.}
\label{fig-pop}
\end{figure}

Fig.~\ref{fig-pop} presents the adiabatic population dynamics obtained from the QD scheme. The CAS adiabatic states with a reference geometries are used as the diabatic states during a propagation segment, which are then dynamically updated for the subsequent propagation steps. The frontier orbitals (HOMO and LUMO) of the on-the-fly CAS(2,2) calculations are visualized along a given trajectory in Fig.~\ref{fig-pop}A. The dynamics are propagated with the PLDM or the SQC approaches, with the results presented in Fig.~\ref{fig-pop}B and Fig.~\ref{fig-pop}C, respectively. For comparison, FSSH with decoherence correction ~\cite{Granucci2007} is also used to generate the photodynamics. For the trajectory-based approaches, a total of 120 trajectories are used to compute the population, with a nuclear time step $dt=0.1$ fs, although a much larger time-step $dt=0.5$ fs can be used and generates the same results at the single trajectory level. The nuclear initial configurations are sampled from the Wigner distribution of the ground vibrational state ($\nu=0$) on the ground electronic state $\mathrm{S}_0$, with the harmonic approximation based on the approach outlined in Ref.~\cite{Wigner}. The electronic DOF (mapping variables in PLDM/SQC and the electronic coefficients in FSSH) is propagated based on the QD scheme with 100 time steps in each nuclear time step. Further numerical details of these calculations are provided in SI. In addition, results obtained from AIMS simulation~\cite{Tao2009jpca} are also presented for comparison. Since AIMS is a wavepacket based approach which has been extensively tested~\cite{Ben-Nun1998jcp,Ben-Nun2000aims,Curchod2018chemrev}, we consider it as an almost exact solution for the quantum dynamics of the ``CAS ethylene model system'', and use it as the benchmark of our calculations. Other recently developed wavepacket approach, such as multiconfigurational Ehrenfest (MCE) method provides essentially the same results as AIMS for this test case at the same level of electronic structure theory~\cite{mcEhrenfest}.

Fig. \ref{fig-pop}B presents the comparison of the population dynamics obtained from QD-PLDM (solid lines) and the decoherence corrected FSSH (dashed lines), and AIMS (filled circles). The population differences between the trajectory based approaches and AIMS are presented in the bottom panel. All three approaches provide the same plateau of the $\mathrm{S}_1$ population ($t=0\sim$20 fs), which corresponds to the initial {\it adiabatic} nuclear relaxation process on the ${\mathrm S}_1$ surface. During $t=20\sim 75$ fs, the system starts to exhibit quick non-adiabatic transitions between ${\mathrm S}_1$ and ${\mathrm S}_0$ states, through conical intersections. Here, QD-PLDM agrees reasonably well with AIMS throughout the entire non-radiative decay process. FSSH, on the other hand, predicts a much faster relaxation dynamics and exhibits a large deviation compare to the AIMS, likely caused by the over-coherence problem despite being correct by a simple decoherence scheme in this calculation. More sophisticated decoherence corrections~\cite{subotnik2016arpc} might further improve the results of FSSH. It is worth noting that the experimentally ~\cite{Tao2011jcp} measured $\mathrm{S}_1$ decay time is $\sim 89$ fs, agrees well with the AIMS when using CASPT2 level of the electronic structure calculations that include dynamical correlation~\cite{Tao2009jpca}. Our intention in this paper, on the other hand, is not trying to compare or recover the experimental results, but rather comparing to the almost exact quantum dynamics of the ``CAS(2,2) ethylene model" provided by AIMS. 

Fig.~\ref{fig-pop}C presents a similar comparison of the population dynamics obtained from QD-SQC (solid lines), the decoherence corrected FSSH (dashed lines), and AIMS (filled circles) with the difference between the trajectory-based approaches and AIMS provided in the bottom panel. Here, we choose to use the simplest possible square window function~\cite{MillerJCP13} proposed by Cotton and Miller. The QD-SQC provides a similar level of accuracy for the population dynamics compared to QD-PLDM, and agrees reasonably well with AIMS throughout the non-radiative decay process. A slightly noisy population is obtained due to the fact that only a fraction of the mapping trajectories landed in all population window at any given time thus reducing the quality of the data and at the same time, requiring normalization of the population~\cite{MillerJCP13}. Nevertheless, QD-SQC still outperforms FSSH in this on-the-fly CAS(2,2) model. We note that more accurate results for model systems can be obtained by using triangle windows~\cite{Cotton2019JCP} and the trajectory specific zero-point energy correction technique~\cite{Cotton2019JCP2}. These new developments will be investigated through the {\it ab-initio} NAMD simulations by using the QD scheme. 

Through results presented here, we demonstrate that the QD scheme enables many possibilities of using recently developed quantum dynamics approaches for accurate {\it ab-initio} on-the-fly NAMD simulation, through the seamless interface between the diabatic quantum dynamics method and the adiabatic electronic structure calculations. On the other hand, the QD propagation scheme also provides new opportunities to assess the performance of approximate diabatic dynamics approaches, with {\it ab-initio} test cases beyond simple diabatic model systems.
\begin{figure*}
\centering
\includegraphics[width=0.9\linewidth]{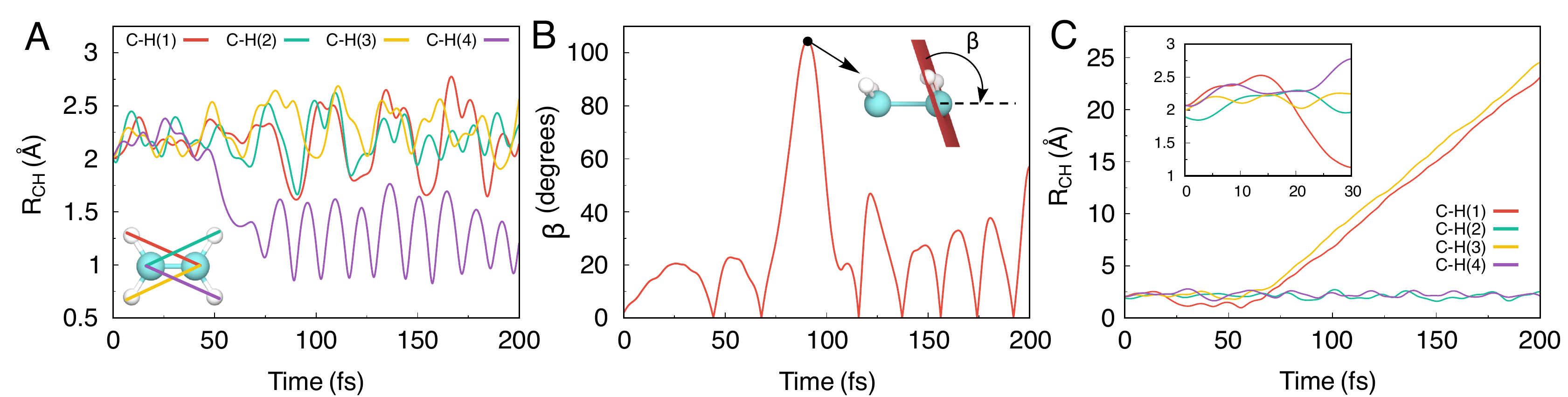}
\caption{Representative QD-PLDM reactive trajectories for (A) hydrogen transfer (B) pyramidalization, and (C) H$_{2}$  dissociation pathways, respectively. The C-H bond lengths are defined in the inset of panel A, and used in both panels A and C. The pyramidalization angle is defined in the inset of panel B.}
\label{fig-traj}
\end{figure*}

\begin{figure}
\centering
\includegraphics[width=0.9\linewidth]{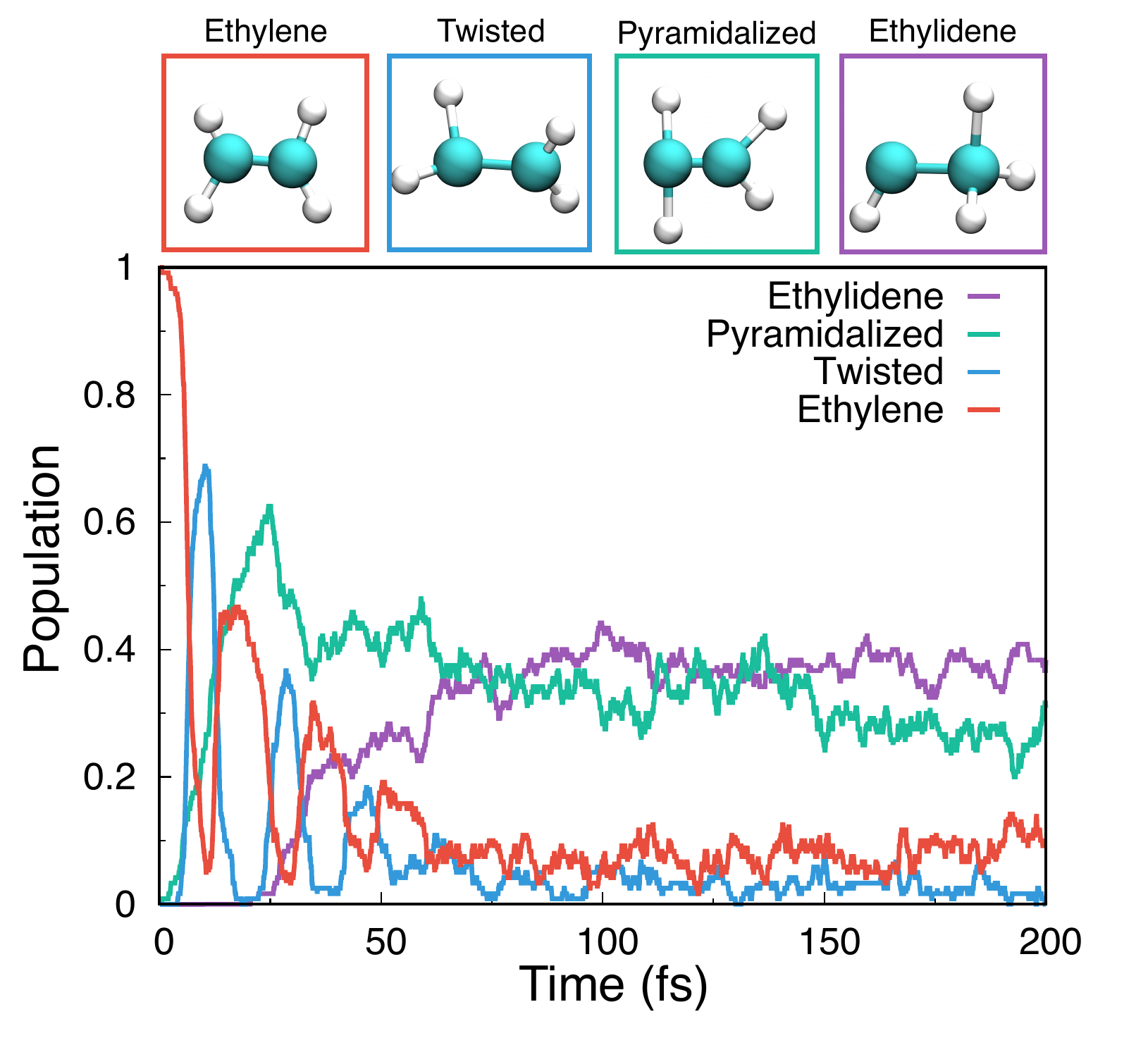}
\caption{Population dynamics of various dissociation products obtained from the QD-PLDM simulations, with representative geometries presented in top panels.}
\label{Product}
\end{figure}

Fig.~\ref{fig-traj} presets three representative reactive trajectories obtained from QD-PLDM, whereas the averaged populations of different nuclear configurations are provided in Fig.~\ref{Product}. A qualitatively similar ensemble of reactive trajectories are also obtained from QD-SQC (not shown). These reactive trajectories provide intuitive time-dependent mechanistic insights into the competing non-radiative decay channels, although a physically meaningful interpretation should only be drawn from the expectation values (such as those presented in Fig.~\ref{Product}). Fig.~\ref{fig-traj}A presents the time-evolution of the bond distance between carbon and hydrogen atoms that are not initially bonded. At $t\approx50$ fs, one of these four distances suddenly drops from  $\sim$2.2 $\AA$ to 1.2 $\AA$, indicating the formation of an ethylidene structure  through the ethylidene-like conical intersection~\cite{Tao2009jpca} (which is different than the twisted-pyramidalized conical intersection shown in Fig.~\ref{fig-pes}). Fig.~\ref{fig-traj}B presents the time evolution of the (modulus of) pyramidalization angle defined in the inset of this panel, forming a persisting oscillation pattern. The zero value of the angle indicates that the molecule going through the planar structure and vibrates on other side of the molecular plane. The inset provides the structure of the largest pyramidalization angle at $\approx 100 ^{\circ}$, which is close to the twisted-pyramidalized conical intersection~\cite{Barbatti2004jcp} shown in Fig.~\ref{fig-pes}. Fig.~\ref{fig-traj}C presents a reactive trajectory of H$_2$ dissociation, which occurs at $t \approx 70$ fs after one H atom abstraction process, as can be seen from the inset of this panel (C-H(1) bond length shrinking indicated by the red curve). These reactive trajectory are in a close agreement with the similar reactive channels discovered from the AIMS simulation~\cite{Levine2008,Tao2009jpca}. 

Fig.~\ref{Product} presents the population of various nuclear configurations obtained from QD-PLDM through the ensemble average of trajectories. These nuclear configurations are defined based on the criteria in Ref.~\cite{Barbatti2005jcp}, with the representative geometries provided on top of this figure (squared with the same color coding used in the population curve). At the short time $t\in[0,20]$ fs, the system evolves adiabatically on the $\mathrm{S}_1$ surface, moving along both the twisted and pyramidalized reaction coordinates, accumulating the population for both configurations. The twisted configuration on $\mathrm{S}_1$ also convert into the pyramidalized configuration during this time. Note that the oscillation period of twisted configuration is around 20 fs, consistent with results obtained from AIMS~\cite{Levine2008} and MCE approach~\cite{mcEhrenfest}. After the early time relaxation on the $\mathrm{S}_1$ surface, the system exhibit various conical intersections and make a non-adiabatic transitions to the $\mathrm{S}_0$ surface, relaxing back to the ethylene configuration (red), or ended up with ethylidene configuration (magenta) or dissociating H$_2$ out of ethylene (with only 8 reactive trajectories out of 120 trajectories, and thus not shown in this figure). Our QD-PLDM simulation predicts that about 50\% of the molecules go through the ethylidene-like conical intersection and the other 50\% of the molecules go through the twisted-pyramidalized conical intersection, agrees well with the AIMS results performed at the CASSCF level of theory~\cite{Levine2008,Tao2009jpca}.

\section*{Concluding Remarks}
In this paper, we provide the first {\it ab-initio} on-the-fly example of using the QD scheme~\cite{Mandal:2018} for non-adiabatic simulation with {\it diabatic} quantum dynamics approach. With two recently developed diabatic dynamics approaches (PLDM and SQC) and on-the-fly CASSCF calculations, we demonstrate the power of the QD scheme by simulating the on-the-fly non-adiabatic dynamics of the ethylene photo-deactivation process. During each short-time propagation segment, the adiabatic states associated with a reference geometry is used as the quasi-diabatic (local diabatic) states, allowing any diabatic dynamics approach to propagate the quantum dynamics during this time step. Between two consecutive propagation segments, the definition of the quasi-diabatic states is updated. The QD scheme thus allows a seamless interface between diabatic dynamics approaches with adiabatic electronic structure calculations, completely eliminates the necessity of any representation reformulating efforts (such as constructing global diabatic or reformulating diabatic dynamics approach to adiabatic representation). It sends out an assurance message to the quantum dynamics community that a diabatic dynamics approach can be directly interfaced with the adiabatic electronic structure calculations to perform on-the-fly simulations. The results obtained from both QD-PLDM and QD-SQC are in close agreement with AIMS; both outperforms the widely used FSSH approach. This work completes the establishment of the QD scheme in the field of {\it ab-initio} non-adiabatic dynamics simulation, demonstrating the QD scheme as a powerful tool to enable accurate diabatic quantum dynamics approaches for on-the-fly simulations.  The QD scheme opens up many possibilities to enable recently developed diabatic dynamics approaches for on-the-fly NAMD simulations, provide alternative theoretical tools compared to the widely used approaches such as FSSH and AIMS. These {\it ab-initio} on-the-fly test cases, on the other hand, provides opportunities to assess the performance of approximate diabatic dynamics approaches beyond simple diabatic model systems, and will foster the development of new quantum dynamics approaches.

\matmethods{\subsection*{Calculation Details} All simulations are performed using a modified version of the SHARC non-adiabatic dynamics interface package~\cite{SHARC,SHARC2}, with the on-the-fly electronic structure calculations  performed with MOLPRO~\cite{MOLPRO-WIREs}. Computational details of the QD-PLDM, QD-SQC, and FSSH, as well as other technical details including system initialization, Wigner sampling, algorithm to track the phase of adiabatic states, and L$\ddot{\mathrm{o}}$wdin orthonormalizations are provided in SI.
\subsection*{\small Potential and Gradient Matrix Elements in the Quasi-Diabatic Representation} During a short-time propagation of the nuclear DOF for $t\in[t_0, t_1]$, the QD scheme uses the adiabatic basis $\{|\Phi_{\alpha}({R}(t_0))\rangle\}$ as the {\it quasi-diabatic} basis. The electronic Hamiltonian operator $\hat V ({R}(t))$ in the QD basis is evaluated as
\begin{equation}\label{eqn:vijt} 
V_{\alpha\beta}({R}(t))  = \langle \Phi_\alpha ({R_0})| \hat V ({R}(t))| \Phi_\beta({R_0})\rangle.
\end{equation}
For on-the-fly simulation, this quantity is obtained from a linear interpolation~\cite{Rossky-Webster} between $V_{\alpha\beta}({R}(t_{0}))$ and $V_{\alpha\beta}({R}(t_{1}))$ as follows
\begin{align}\label{eqn:interpolation}
V_{\alpha\beta}({R}(t)) & = V_{\alpha\beta}(R_{0}))+\frac {(t - t_{0})}{(t_{1} - t_{0})}\big[V_{\alpha\beta}({R}(t_{1})) - V_{\alpha\beta}({R}_{0}))\big],
\end{align} 
where $V_{\alpha\beta}({R}(t_{0})) =\langle \Phi_\alpha ({R_0})| \hat V ({R}(t_0))| \Phi_\beta ({R_0})\rangle = E_{\alpha}({R}(t_0))\delta_{\alpha\beta}$. The matrix elements $V_{\alpha\beta}({R}(t_{1}))$ are computed as follows
\begin{equation}\label{eqn:elect2}
V_{\alpha\beta}({R}(t_{1})) =\sum_{\lambda\nu}S_{\alpha\lambda} {V}_{\lambda\nu} ({R}(t_1)) S^{\dagger}_{\beta\nu},   
 \end{equation}
where ${V}_{\lambda\nu} ({R}(t_1))=\langle \Phi_{\lambda}({R}(t_{1}))| \hat {V} ({R}(t_1))|\Phi_{\nu}({R}(t_{1})) \rangle=E_{\lambda}({R}(t_1))\delta_{\lambda\nu}$, and the overlap matrix between two adiabatic electronic states (with two different nuclear geometries) are  $S_{\alpha\lambda}= \langle \Phi_{\alpha}({R_0})|\Phi_{\lambda}({R}(t_{1}))\rangle$ and $S^{\dagger}_{\beta\nu} = \langle \Phi_{\nu}({R}(t_{1}))|\Phi_{\beta}({R_0})\rangle$. These overlap matrix are computed based on the approach outlined in Ref.~\cite{Plasser2016JCTC}.

The nuclear gradients $\nabla V_{\alpha\beta}({R}(t_{1}))\equiv \partial V_{\alpha\beta}({R}(t_{1}))/\partial {R}$ are evaluated as
\begin{eqnarray}\label{eqn:nucgrad}
\nabla V_{\alpha\beta}({R}(t_{1}))=\sum_{\lambda\nu} S_{\alpha\lambda}\langle \Phi_{\lambda}({R}(t_{1}))|\nabla \hat V ({R}(t_1))|\Phi_{\nu}({R}(t_{1})) \rangle S^{\dagger}_{\beta\nu}.
\end{eqnarray}  
We emphasize that Eqn.~\ref{eqn:nucgrad} includes derivatives with respect to all possible sources of the nuclear dependence, including those from the adiabatic potentials as well as the adiabatic states~\cite{Mandal2018jcp,Mandal2019jpca}. The details of this justification is provided in the SI.

During the next short-time propagation segment $t\in[t_1,t_2]$, the QD scheme adapts a new reference geometry ${R'_{0}}\equiv {R}(t_1)$ and new {\it diabatic} basis $|\Phi_{\mu}({R'_{0}})\rangle\equiv|\Phi_{\mu}({R}(t_1))\rangle$. Between $[t_0,t_1]$ propagation and $[t_1,t_2]$ propagation segments, all of these quantities will be transformed from $\{|\Phi_{\alpha}({R_{0}})\rangle\}$ to $\{|\Phi_{\mu}({R'_{0}})\rangle\}$ basis, using the relation 
\begin{equation}\label{eqn:basis}
|\Phi_{\lambda}({R}(t_{1}))\rangle=\sum_{\alpha} \langle \Phi_{\alpha}({R}(t_{0}))| \Phi_{\lambda}({R}(t_{1}))\rangle|\Phi_{\alpha}({R}(t_{0}))\rangle.
\end{equation}

Note that the QD propagation scheme does not explicitly require the derivative couplings ${d}_{\lambda\nu}({R})=\langle \Phi_{\lambda}({R})|\nabla \Phi_{\nu}({R})\rangle$ or non-adiabatic coupling $\langle \Phi_\beta({ R}(t))|{\partial\over{\partial t}}\Phi_\alpha({ R}(t))\rangle={ d}_{\beta\alpha}({R})\dot{R}$. That said, the QD scheme does not omit these quantities either; the nuclear gradient $\nabla V_{\alpha\beta}({R}(t_{1}))$ now contains $\langle \Phi_{\lambda}({R}(t_{1}))|\nabla \hat{V} ({R}(t_1))|\Phi_{\nu}({R}(t_{1}))\rangle$ (see Eqn.~\ref{eqn:nucgrad}), which is reminiscent of the derivative coupling, and the QD scheme uses transformation matrix elements $\langle\Phi_\beta({ R}(t_1))|\Phi_\alpha({ R}(t_2))\rangle$ instead of $\langle \Phi_\beta({R}(t))|{\partial\over{\partial t}}\Phi_\alpha({R}(t))\rangle$. It is worth noting that both ${d}_{\lambda\nu}({R})$ and $\langle \Phi_\beta({ R}(t))|{\partial\over{\partial t}}\Phi_\alpha({ R}(t))\rangle$ can become singular. The QD scheme explicitly alleviates this difficulty by using the well behaved quantities $\nabla V_{\alpha\beta}({R}(t_{1}))$ and $\langle\Phi_\beta({ R}(t_1))|\Phi_\alpha({ R}(t_2))\rangle$. Thus, a method that directly requires derivative couplings and/or non-adiabatic coupling might suffer from numerical instabilities near trivial crossings or conical intersections, whereas a method that only requires the gradient (such as the QD scheme) will likely not~\cite{Sandoval2018}. 

\subsection*{Partial Linearized Density Matrix (PLDM) Path-Integral Approach}
PLDM is an approximate quantum dynamics method based on the real-time path-integral approach~\cite{HuoPLDM}. Using the MMST mapping representation,\cite{StockThossPRA99} the non-adiabatic transitions among discrete electronic states $\{|i\rangle, |j\rangle\}$ are exactly mapped~\cite{StockThossPRA99} onto the phase-space motion of the fictitious variables through the relation $|i \rangle\langle j | \rightarrow  {\hat a}_{i}^\dagger {\hat a}_{j}$, where ${\hat a}^\dagger_{i} =({\hat q}_{i} - i{\hat p}_{i})/\sqrt{2}$ and ${\hat a}_{i} =({\hat q}_{i} + i{\hat p}_{i})/\sqrt{2}$. After performing the linearization approximaiton on the nuclear DOF, we obtain the following PLDM reduced density matrix~\cite{HuoPLDM} 
\begin{equation}\label{eq:redmat}
\rho_{ij}(t)=\mathrm{Tr}_{R}[\hat{\rho}(0)e^{i\hat{H}t/\hbar}|i\rangle\langle j|e^{-i\hat{H}t/\hbar}]\approx \sum_{kl}\int d\tau  [\hat{\rho}(0)^\mathrm{W}_{kl}]  T_{ki}(t) T'_{jl}(t),
\end{equation}
where $\int d\tau  \equiv \frac{1}{2\pi\hbar}\int d{R} dP d{\bf q} d{\bf p} d{\bf q'} d{\bf p'} G_0G'_0 $ represents the phase space integration for all DOFs with $G_0$ and $G'_0$ represents coherence state distribution of mapping oscillators, $T_{ki}(t)=\frac{1}{2}(q_{i}(t)+ip_{i}(t))(q_{k}(0)-ip_{k}(0))$ and $T'_{jl}(t)=\frac{1}{2}(q'_{l}(0)+ip'_{l}(0))(q'_{j}(t)-ip'_{j}(t))$ are the electronic transition amplitudes associated with the forward mapping trajectory $\{{\bf q}, {\bf p}\}$ and the backward mapping trajectory $\{{\bf q'}, {\bf p'}\}$, respectively. $[\hat{\rho}(0)^\mathrm{W}_{kl}]$ is the partial Wigner transform (with respect to the nuclear DOF) of the $kl_\mathrm{th}$ matrix element of the initial total density operators $\hat{\rho}(0)$. 

Classical trajectories are used to evaluate the approximate time-dependent reduced density matrix. The forward mapping variables are evolved based on the Hamilton's equations of motion~\cite{HuoPLDM,Huo2016ARPC} $\dot q_{i} = \partial H_\mathrm{m} / \partial p_{i}$ and $\dot p_{i} = -\partial H_\mathrm{m} / \partial q_{i}$, where $H_\mathrm{m}={{{P}^2}\over {2M}}+{1\over2}\sum_{ij}V_{ij}(R)\big[p_{i}p_{j}+q_{i}q_{j}\big]$ is the PLDM mapping Hamiltonian~\cite{HuoPLDM}. The backward mapping variables are propagated with the similar equations of motion governed by $H_\mathrm{m}({\bf p'},{\bf q'})$. The nuclei are evolved with the force ${F}=-{1\over 4}\sum_{ij}\nabla V_{ij}(R)[p_{i}p_{j}+q_{i}q_{j}+p'_{i}p'_{j}+q'_{i}q'_{j}]$. All of the necessary elements are evaluated with the QD scheme outlined above, with the technical details provided in SI.
}
\showmatmethods{} 

\acknow{This work was supported by the National Science Foundation CAREER Award under Grant No. CHE-1845747 as well as by ``Enabling Quantum Leap in Chemistry'' program under a Grant number CHE-1836546. W. Z. appreciates the support from China Scholarship Council (CSC). A. M. appreciate the support from the Elon Huntington Hooker Fellowship. Computing resources were provided by the Center for Integrated Research Computing (CIRC) at the University of Rochester. W.Z. appreciates valuable discussions with Daniel Hollas. We appreciate the generous technical support and valuable and timely feedback from the SHARC developer team by Dr. Sebastian Mai and Prof. Leticia Gonz$\grave{a}$lez. P.H. appreciates valuable discussions with Profs. Ben Levine, Artur Izmaylov, and Dmitry Shalashilin, as well as Prof. Bill Miller for his encouragement that initiated this work. }
\showacknow{}


\end{document}